\documentclass{article}
\usepackage{amssymb}

\usepackage{graphicx}
\usepackage{amsmath}

\begin{document}

\textbf{Classical and Quantum Interaction of the Dipole Revisited}\medskip
\bigskip

\qquad Tomislav Ivezi\'{c}

\qquad\textit{Ru%
\mbox
{\it{d}\hspace{-.15em}\rule[1.25ex]{.2em}{.04ex}\hspace{-.05em}}er Bo\v
{s}kovi\'{c} Institute, P.O.B. 180, 10002 Zagreb, Croatia}

\textit{\qquad ivezic@irb.hr\bigskip \medskip }

The interaction of the electric and magnetic dipole moments of a particle
with the electromagnetic field is investigated in an approach that deals
with four-dimensional (4D) geometric quantities. The new commutation
relations for the 4D orbital and intrinsic angular momentums and also for
the 4D dipole moments are introduced. The expectation value of the quantum
4-force, which holds in any frame, is worked out in terms of them. In
contrast to it the whole calculation in [1] ([1] J. Anandan, Phys. Rev.
Lett. \textbf{85}, 1354 (2000)) has been made only in the rest frame of the
dipole. It is proved that, e.g., the expression for the 3D force $\mathbf{f}%
_{S}$ in [1] is not relativistically correct and that the quantum 4-force is
not zero in the experiments proposed in [1]. This means that the phase
shifts that could be observed in such experiments are not topological phase
shifts.\bigskip \medskip

\bigskip

\medskip \bigskip

\noindent \textbf{1. Introduction}\bigskip

In a recent paper [1] Anandan has presented a covariant treatment of the
interaction of the electric and magnetic dipole moments of a particle with
the electromagnetic field. In [2] some objections to such treatment have
been raised. A geometric approach to special relativity is considered in [2]
in which a physical reality is attributed to the four-dimensional (4D)
geometric quantities and not, as usually accepted, to the 3D quantities. The
same approach and the results from [2], and [3], will be used in this paper.
Instead of the interaction term, Eq. (5) in [1], much more general
expression has been derived in [2], with tensors as 4D geometric quantities.
In [1] Anandan also calculated the expectation value of the quantum force
operator for two models; the dipole which is made of two particles with
charges $q$, $-q$ and an elementary particle with its intrinsic magnetic and
electric dipole moments. The whole calculation is made only in the rest
frame of the dipole. Here, a more general expression for the force from [1]
(for the second model), which holds in any frame, will be presented using
the geometric approach. The 4-force is calculated by means of the new
commutation relations for the orbital (1) and intrinsic angular momentum
4-vectors and also for the dipole moments (2). This force is compared with
the force $\mathbf{f}_{S}$, Eq. (26) in [1]. It is found that $\mathbf{f}%
_{S} $ from [1] is not relativistically correct and that, contrary to the
assertion from [1], the quantum 4-force is not zero in the experiments
proposed in [1].\bigskip

\noindent \textbf{2. 4D geometric approach}\bigskip

In the same way as in [2], [3] and [4], the 4D geometric quantities, that
are defined without reference frames, i.e., the absolute quantities (AQs),
e.g., the 4-vectors of the electric and magnetic fields $E^{a}$ and $B^{a}$,
the electromagnetic field tensor $F^{ab}$, the dipole moment tensor $D^{ab}$%
, the 4-vectors of the electric dipole moment (EDM) $d^{a}$ and the magnetic
dipole moment (MDM) $m^{a}$, the spin four-tensor $S^{ab}$, etc. will be
employed here. Also, we shall deal with the representations of the AQs in
the standard basis. Note that usually, including [1], only the component
form of tensors in the standard basis is used. (The standard basis \{$e_{\mu
};\ 0,1,2,3$\} consists of orthonormal 4-vectors with $e_{0}$ in the forward
light cone. It corresponds to the specific system of coordinates with
Einstein's synchronization [5] of distant clocks and Cartesian space
coordinates $x^{i}$.)

Let us start quoting some results from [2], which are used in [3] and [4] as
well. First, $F^{ab}$, as the primary quantity for the whole
electromagnetism [6], can be decomposed into $E^{a}$, $B^{a}$ and the
4-velocity $v^{a}$ of the observers who measure fields; $%
F^{ab}=(1/c)(E^{a}v^{b}-E^{b}v^{a})+\varepsilon ^{abcd}v_{c}B_{d}$, whence $%
E^{a}$, $B^{a}$ are derived as $E^{a}=(1/c)F^{ab}v_{b}$ and $%
B^{a}=(1/2c^{2})\varepsilon ^{abcd}F_{bc}v_{d}$, with $%
E^{a}v_{a}=B^{a}v_{a}=0$. The frame of ``fiducial'' observers, in which the
observers who measure $E^{a}$, $B^{a}$ are at rest with the standard basis $%
\{e_{\mu }\}$ in it is called the $e_{0}$-frame. In the $e_{0}$-frame $%
E^{0}=B^{0}=0$ and $E^{i}=F^{i0}$, $B^{i}=(1/2c)\varepsilon ^{ijk0}F_{jk}$.
The similar relations hold for $D^{ab}$, $d^{a}$ and $m^{a}$, and the
4-velocity of the particle $u^{a}$, $u^{a}=dx^{a}/d\tau $. Then $%
D^{ab}=(1/c)(u^{a}d^{b}-u^{b}d^{a})+(1/c^{2})\varepsilon ^{abcd}u_{c}m_{d}$,
$m^{a}=(1/2)\varepsilon ^{abcd}D_{bc}u_{d}$, $d^{a}=(1/c)D^{ba}u_{b}$, with $%
d^{a}u_{a}=m^{a}u_{a}=0$. Only in the particle's rest frame (the $K^{\prime
} $ frame) and the $\{e_{\mu }^{\prime }\}$ basis $d^{\prime 0}=m^{\prime
0}=0$, $d^{\prime i}=D^{\prime 0i}$, $m^{\prime i}=(c/2)\varepsilon
^{ijk0}D_{jk}^{\prime }$. Observe that in [1] only one velocity $u^{a}$ is
used both for the determination of $d^{a}$, $m^{a}$ from $D^{ab}$ and the
determination of $E^{a}$, $B^{a}$ from $F^{ab}$. (Actually in [1] only
components are considered and the decompositions of $F^{ab}$ and $D^{ab}$
are never used.) This is objected in [2], where different 4-velocities $%
v^{a} $ and $u^{a}$ are introduced. The Eqs. (7) and (6) from [1] become $%
(1/2)F_{ab}D^{ba}=(1/c)D_{a}u^{a}+(1/c^{2})M_{a}u^{a}=a^{b}u_{b}$, where $%
a_{b}=(1/c)D_{b}+(1/c^{2})M_{b}$, and $D_{b}=d^{a}F_{ab}$, $%
M_{b}=m^{a}F_{ab}^{\ast }=(1/2)m^{a}\varepsilon _{abcd}F^{cd}$. Hence Eq.
(5) from [1] becomes more complicated. $(1/2)F_{ab}D^{ba}$ is written as the
sum of two terms $%
(1/c^{2})[((E_{a}d^{a})+(B_{a}m^{a}))(v_{b}u^{b})-(E_{a}u^{a})(v_{b}d^{b})-(B_{a}u^{a})(v_{b}m^{b})]
$ and $(1/c^{3})[\varepsilon
^{abcd}(v_{a}E_{b}u_{c}m_{d}+c^{2}d_{a}u_{b}v_{c}B_{d})]$. As seen from the
second term it naturally contains the interaction of $E^{a}$ with $m^{a}$,
and $B^{a}$ with $d^{a}$. Such interactions are required for the
explanations of the Aharonov-Casher effect and the R\"{o}ntgen phase shift,
as shown in [2,4], and also of different methods of measuring EDMs, e.g.,
[7], which is considered in [3]. Note that for the comparison with
experiments we only need to choose the laboratory frame as our $e_{0}$-frame
and then to represent the AQs $E^{a}$, $m^{a}$ and $B^{a}$, $d^{a}$ in that
frame.\bigskip

\noindent \textbf{3. The quantum phase shift }\bigskip

The quantum phase shift that the particle experiences due to the field is
given by Eqs. (9), (10) and (11) in [1], but with our $a_{\mu }$;
\begin{equation}
a_{\mu }=(1/c)d^{\nu }F_{\nu \mu }+(1/2c^{2})m^{\nu }\varepsilon _{\nu \mu
\rho \lambda }F^{\rho \lambda }.  \label{a}
\end{equation}
As said in [1] $d^{\nu }$ and $m^{\nu }$, and therefore $a_{\mu }$ are now
operators which need not commute. ($E^{\nu }$ and $B^{\nu }$ are not
operators; they are not quantized fields.) In [1] $a_{\mu }$ is given by Eq.
(16). The components $a_{0}$ and $a_{i}$ in Eq. (16) in [1] are written in
terms of $\mathbf{E}$, $\mathbf{B}$, $\mathbf{d}$ and $\mathbf{m}$ and it is
stated that Eq. (16) is a low energy approximation. (The vectors in the 3D
space will be designated in bold-face.) Our $a_{\mu }$ is obtained inserting
the decomposition of $F^{\mu \nu }$ ($F^{\mu \nu }=(1/c)(E^{\mu }v^{\nu
}-E^{\nu }v^{\mu })+\varepsilon ^{\mu \nu \rho \lambda }v_{\rho }B_{\lambda
} $) into $a_{\mu }$ and it differs in several important respects relative
to $a_{\mu }$ from [1]. First, we always deal with 4D quantities and not
with the 3D vectors. Furthermore, we do not need to make a low energy
approximation. The comparison with [1] can be made writing our $a_{\mu }$ in
the $e_{0}$-frame. There, $v^{\mu }=(c,0,0,0)$ and $E^{0}=B^{0}=0$. Hence
\begin{eqnarray}
a_{0} &=&(-1/c)(d^{i}E^{i}+m^{i}B^{i}),  \notag \\
a_{i} &=&(1/c)(d^{0}E^{i}+m^{0}B^{i})+\varepsilon
^{0ijk}d^{j}B^{k}-(1/c^{2})\varepsilon ^{0ijk}m^{j}E^{k};  \label{a0}
\end{eqnarray}
the metric is diag$(1,-1,-1,-1)$, $\varepsilon ^{0123}=1$ and the components
of the 3D vectors correspond to the components of the 4D vectors with upper
indices. Since the experiments are made in the laboratory frame (the $K$
frame) we shall choose that $K$ is our $e_{0}$-frame. As already said, in
[1] only the particle's 4-velocity is considered and all 3D vectors are
written in $K^{\prime }$, the particle's rest frame. Hence, only when $%
K^{\prime }$ is the $e_{0}$-frame then Eq. (16) from [1] is recovered, but
with the components of the 4D quantities. However that case is physically
unrealizable since in the experiments the observers do not ``seat'' on the
particle. As seen from [2] this ambiguity of the theory from [1] is simply
avoided in our formulation with two 4D geometric velocities $v^{a}$ and $%
u^{a}$. Furthermore, as explained in [8], [9] and [2], the transformations
of $\mathbf{E}$ and $\mathbf{B}$ (e.g., [10] Eq. (11.149)), and also of $%
\mathbf{d}$ and $\mathbf{m}$, are all the ``apparent'' transformations (AT)
of the 3D vectors and not the Lorentz transformations (LT). The AT do not
refer to the same 4D quantity and, [8], [9] and [2], they are not
relativistically correct transformations. Therefore, contrary to the
assertion from [1], Eq. (16) cannot be transformed in a relativistically
correct way to the laboratory frame. For our $a_{\mu }$ it will hold that $%
a^{b}=a^{\mu }e_{\mu }=a^{\prime \mu }e_{\mu }^{\prime }$, as for all other
4D geometric quantities; it is the same quantity for relatively moving
observers in $K$ and $K^{\prime }$. Observe that all primed quantities are
the Lorentz transforms of the unprimed ones. On the other hand such relation
does not hold for the quantities from [1], e.g., $a^{\prime \mu }$ Eq. (16),
$\mathbf{f}^{\prime }$ Eq. (25) and $\mathbf{f}_{S}^{\prime }$ Eq. (26),
since all 3D vectors transform by the AT and not by the LT.\bigskip

\noindent \textbf{4. The quantum 4-force, I }\bigskip

In [1] the 4-force (in fact, the components in the $\{e_{\mu }\}$ basis) is
defined by Eq. (12), which we write as $f^{\mu }\equiv m\overset{\cdot \cdot
}{\xi }^{\mu }=\left\langle \psi \mid G_{\ \nu }^{\mu }\mid \psi
\right\rangle \overset{\cdot }{\xi }^{\nu }$, where $\xi ^{\mu }$ denotes
the expectation value of the components of the 4-position operator, the dot
denotes the derivation over the proper time of the particle, and $G_{\mu \nu
}=\partial _{\mu }a_{\nu }-\partial _{\nu }a_{\mu }-(i/$%
h\hskip-.2em\llap{\protect\rule[1.1ex]{.325em}{.1ex}}\hskip.2em%
$)[a_{\mu },a_{\nu }]$, Eq. (11) in [1]. In our calculation the 4-force, as
a geometric quantity, will be written as $f^{\mu }e_{\mu }=\left\langle \psi
\mid G_{\ \nu }^{\mu }\mid \psi \right\rangle u^{\nu }e_{\mu }$, where, from
now on, the expectation value $\overset{\cdot }{\xi }^{\nu }$ is denoted as $%
u^{\nu }$. In [1], the forces $\mathbf{f}^{\prime }$ Eq. (25) and $\mathbf{f}%
_{S}^{\prime }$ Eq. (26) are determined in the rest frame of the particle
(i.e., the dipole) where $f^{\prime 0}=0$ (in our notation all quantities in
Eqs. (25) and (26) are the primed quantities). There, in [1], it is argued
``The force in an arbitrary inertial frame may be obtained by Lorentz
transforming the above $f^{\mu }$ (Eq. (25), our remark) to this frame.''
But, as discussed above, this statement is not true since the
transformations of 3D quantities from Eqs. (25) and (26) are the AT and not
the LT. Therefore we shall derive the expression for $f_{S}^{\mu }e_{\mu }$
for an elementary particle that holds in any inertial frame.

When deriving $\mathbf{f}_{S}$, Eq. (26) in [1], it is asserted: ``For an
elementary particle, the only intrinsic direction is provided by the spin $%
\mathbf{S}$. Then its intrinsic $\mathbf{\mu }=\gamma _{S}\mathbf{S}$ and
its intrinsic $\mathbf{d}=\delta _{S}\mathbf{S}$, where $\delta _{S}$ is a
constant.'' (In [1], as already said, the unprimed quantities are in the
particle's rest frame $K^{\prime }$.) Thus both the 3D MDM $\mathbf{m}%
^{\prime }$ and the 3D EDM $\mathbf{d}^{\prime }$ (our notation) of an
elementary particle are determined by the usual 3D spin $\mathbf{S}^{\prime
} $. Then, only the commutation relation for the components of the 3D spin
operator ($[S^{i},S^{j}]=i$%
h\hskip-.2em\llap{\protect\rule[1.1ex]{.325em}{.1ex}}\hskip.2em%
$\varepsilon ^{ijk}S_{k}$) are used in the calculation of the commutator $%
[a_{i},a_{0}]$, i.e., the commutators of $\mathbf{m}^{\prime }$, $\mathbf{d}%
^{\prime }$ with $\mathbf{m}^{\prime }$, $\mathbf{d}^{\prime }$.

Recently, [9] and [3], it is shown that the angular momentum four-tensor $%
M^{ab}$, $M^{ab}=x^{a}p^{b}-x^{b}p^{a}$, can be decomposed into the
``space-space'' angular momentum of the particle $L^{a}$ and the
``time-space'' angular momentum $K^{a}$ (both with respect to the observer
with velocity $v^{a}$). In [3] a similar consideration is applied to the
\emph{intrinsic} angular momentum, the spin of an elementary particle. The
primary quantity with the definite physical reality is considered to be the
spin four-tensor $S^{ab}$, which is decomposed into two 4-vectors, the usual
``space-space'' intrinsic angular momentum $S^{a}$ and the ``time-space''
intrinsic angular momentum $Z^{a}$, see [3]. Thus, [3] introduces a new
``time-space'' spin $Z^{a}$, which is a physical quantity in the same
measure as it is the usual ``space-space'' spin $S^{a}$. In all usual
approaches only $\mathbf{L}$ is considered to be a well-defined physical
quantity, whose components transform according to the AT, see, e.g., Eq.
(11) in [11], $L_{x}=L_{x}^{\prime }$, $L_{y}=\gamma (L_{y}^{\prime }-\beta
K_{z}^{\prime })$, $L_{z}=\gamma (L_{z}^{\prime }+\beta K_{y}^{\prime })$;
the transformed components $L_{i}$ are expressed by the mixture of
components $L_{k}^{\prime }$ and $K_{k}^{\prime }$. (These transformations
are the same as are the AT for the components of $\mathbf{B}$, e.g., [10]
Eq. (11.148), since $\mathbf{L}$ correspond to $-\mathbf{B}$ and $\mathbf{K}$
correspond to $-\mathbf{E}$.) In our geometric approach a physical reality
is attributed to the whole $M^{ab}$ ($S^{ab}$) or, equivalently, to the
angular momentums $L^{a}$ ($S^{a}$) and $K^{a}$ ($Z^{a}$), which contain the
same physical information as $M^{ab}$ ($S^{ab}$) only when they are taken
together. The components $L^{\mu }$, or $K^{\mu }$, transform by the LT
again to the components $L^{\prime \mu }$ ($L^{\prime 0}=\gamma (L^{0}-\beta
L^{1})$, $L^{\prime 1}=\gamma (L^{1}-\beta L^{0})$, $L^{\prime 2,3}=L^{2,3}$%
, for the boost in the $x^{1}$ - direction), and the same holds for $S^{\mu
} $, or $Z^{\mu }$. Furthermore, in [3], an essentially new connection
between dipole moments and the spin is formulated in terms of the
corresponding 4D geometric quantities as $m^{a}=\gamma _{S}S^{a}$, $%
d^{a}=\delta _{Z}Z^{a}$, where $\gamma _{S}$ and $\delta _{Z}$ are
constants. In the particle's rest frame and the $\{e_{\mu }^{\prime }\}$
basis $d^{\prime 0}=m^{\prime 0}=0$, $d^{\prime i}=\delta _{Z}Z^{\prime i}$,
$m^{\prime i}=\gamma _{S}S^{\prime i}$. Thus, [3], the intrinsic MDM $m^{a}$
of an elementary particle is determined by the ``space-space'' intrinsic
angular momentum $S^{a}$, while the intrinsic EDM $d^{a}$ is determined by
the ``time-space'' intrinsic angular momentum $Z^{a}$.\bigskip

\noindent \textbf{5. New commutation relations }\bigskip

Hence, in the calculation of the commutator $[a_{\mu },a_{\nu }]$ the usual
commutation relations will be generalized taking into account the results
from [3]. From the Lie algebra of the Poincar\'{e} group we know that

\noindent $\lbrack M^{\mu \nu },M^{\rho \sigma }]=-i$%
h\hskip-.2em\llap{\protect\rule[1.1ex]{.325em}{.1ex}}\hskip.2em%
$(-g^{\nu \rho }M^{\mu \sigma }+g^{\mu \rho }M^{\nu \sigma }+g^{\mu \sigma
}M^{\nu \rho }-g^{\nu \sigma }M^{\mu \rho })$. Then, one has to take into
account the decomposition of the components $M^{\mu \nu }$ into $L^{\mu }$
and $K^{\mu }$ (they are now operators), $M^{\mu \nu }=(1/c)[(v^{\mu }K^{\nu
}-v^{\nu }K^{\mu })+\varepsilon ^{\mu \nu \rho \sigma }L_{\rho }v_{\sigma }]$%
, where, for a macroscopic observer, $v^{\mu }$ can be taken as the
classical velocity of the observer (the components), i.e., not the operator.
This leads to the new commutation relations
\begin{eqnarray}
\lbrack L^{\mu },L^{\nu }] &=&(i\hslash /c)\varepsilon ^{\mu \nu \alpha
\beta }L_{a}v_{\beta },\ [K^{\mu },K^{\nu }]=(-i\hslash /c)\varepsilon ^{\mu
\nu \alpha \beta }L_{a}v_{\beta },  \notag \\
\lbrack L^{\mu },K^{\nu }] &=&(i\hslash /c)\varepsilon ^{\mu \nu \alpha
\beta }K_{a}v_{\beta },  \label{i}
\end{eqnarray}
which, in the $e_{0}$-frame, where $L^{0}=K^{0}=0$, reduce to the usual
commutators for the components of $\mathbf{L}$ and $\mathbf{K}$ (as
operators), see, e.g., [12] Eqs. (2.4.18) - (2.4.20). The same commutators
as in (\ref{i}) hold for the intrinsic angular momentums (the components) $%
S^{\mu }$ and $Z^{\mu }$; $S^{\mu }$ replaces $L^{\mu }$, $Z^{\mu }$
replaces $K^{\mu }$ and the velocity of the particle (the components) $%
u^{\mu }$, i.e., the expectation value $\overset{\cdot }{\xi }^{\mu }$,
replaces the velocity of the observer $v^{\mu }$,
\begin{eqnarray}
\lbrack S^{\mu },S^{\nu }] &=&(i\hslash /c)\varepsilon ^{\mu \nu \alpha
\beta }S_{a}u_{\beta },\ [Z^{\mu },Z^{\nu }]=(-i\hslash /c)\varepsilon ^{\mu
\nu \alpha \beta }S_{a}u_{\beta },  \notag \\
\lbrack S^{\mu },Z^{\nu }] &=&(i\hslash /c)\varepsilon ^{\mu \nu \alpha
\beta }Z_{a}u_{\beta }.  \label{sz}
\end{eqnarray}
Note that in [1] only the commutators $[L_{i},L_{j}]$ and $[S_{i},S_{j}]$
appear. Taking into account the relations $m^{\mu }=\gamma _{S}S^{\mu }$ and
$d^{\mu }=\delta _{Z}Z^{\mu }$ one can express the commutation relations for
$m^{\mu }$ and $d^{\mu }$ in terms of those for $S^{\mu }$ and $Z^{\mu }$
\begin{equation}
\lbrack m^{\mu },m^{\nu }]=\gamma _{S}^{2}[S^{\mu },S^{\nu }],\ [d^{\mu
},d^{\nu }]=\delta _{Z}^{2}[Z^{\mu },Z^{\nu }],\ [m^{\mu },d^{\nu }]=\gamma
_{S}\delta _{Z}[S^{\mu },Z^{\nu }].  \label{1}
\end{equation}
\bigskip

\noindent \textbf{6. The quantum 4-force, II }\bigskip

Then, the 4-force (components) $f_{S}^{\mu }$ can be calculated using (\ref
{1}) and (\ref{sz}). The obtained expressions for $f_{S}^{\mu }$ are much
more general but also much more complicated than those in [1]. First we
consider the terms which come from $\partial _{\mu }a_{\nu }-\partial _{\nu
}a_{\mu }$ in $G_{\mu \nu }$. This term

\noindent $(1/c)\left\langle \psi \mid \partial _{\mu }[d^{\alpha
}(E_{\alpha }v_{\nu }-E_{\nu }v_{\alpha })]-\partial _{\nu }[d^{\alpha
}(E_{\alpha }v_{\mu }-E_{\mu }v_{\alpha })]\mid \psi \right\rangle u^{\nu
}g^{\lambda \mu }e_{\lambda }$

\noindent in $f_{S}^{\lambda }e_{\lambda }$ will correspond to $\nabla
^{\prime }(\mathbf{d}^{\prime }\cdot \mathbf{E}^{\prime })$ in $\mathbf{f}%
_{S}^{\prime }$ (our notation), Eq. (26) in [1], when $K^{\prime }$ is taken
to be the $e_{0}$-frame, i.e., when $u^{\mu }=v^{\mu }=(c,0,0,0)$. Similarly
the term which will correspond to $\nabla ^{\prime }(\mathbf{m}^{\prime
}\cdot \mathbf{B}^{\prime })$ in $\mathbf{f}_{S}^{\prime }$ is obtained from
the above quoted term replacing $d^{\alpha }$, $E^{\alpha }$ with $m^{\alpha
}$, $B^{\alpha }$. The other two terms in $\partial _{\mu }a_{\nu }-\partial
_{\nu }a_{\mu }$ are $(1/c)\{\partial _{\mu }[\varepsilon _{\alpha \nu
\sigma \rho }d^{\alpha }v^{\sigma }B^{\rho }]-\mu \leftrightarrows \nu \}$
and $(1/2c^{3})\{\partial _{\mu }[\varepsilon _{\alpha \nu \sigma \rho
}m^{\alpha }(E^{\sigma }v^{\rho }-E^{\rho }v^{\sigma })]-\mu
\leftrightarrows \nu \}$ which, when $K^{\prime }$ is the $e_{0}$-frame,
correspond to the third and the fourth term, respectively, in Eq. (26) in
[1].

The commutator $[a_{\mu },a_{\nu }]$ will give the additional sixteen terms
in our $f_{S}^{\lambda }e_{\lambda }$, which are easily determined using (%
\ref{1}) and the commutation relations for $S^{\mu }$ and $Z^{\mu }$. In the
case when $K^{\prime }$ is the $e_{0}$-frame then only four terms remain.
From the following part $-(i/\hslash )[a_{\mu },a_{\nu }]$ of $G_{\mu \nu }$
we find the general expressions for the mentioned four terms. The two terms
are: $(-\gamma _{S}/2c^{6})\varepsilon _{\alpha \mu \sigma \rho }\varepsilon
^{\alpha \beta \gamma \delta }(E^{\sigma }v^{\rho }-E^{\rho }v^{\sigma
})(B_{\nu }v_{\beta }-B_{\beta }v_{\nu })m_{\gamma }u_{\delta }$, and $%
(\gamma _{S}/c^{6})\varepsilon _{\alpha \mu \sigma \rho }\varepsilon
^{\alpha \beta \gamma \delta }(E_{\beta }E^{\sigma }v_{\nu }v^{\rho }+E_{\nu
}E^{\rho }v_{\beta }v^{\sigma })d_{\gamma }u_{\delta }$. They determine two
terms in $f_{S}^{\lambda }e_{\lambda }$, which correspond to $(-\gamma
_{S}/c^{2})(\mathbf{m}^{\prime }\times \mathbf{B}^{\prime })\times \mathbf{E}%
^{\prime }$ and $(-\gamma _{S}/c^{2})(\mathbf{d}^{\prime }\times \mathbf{E}%
^{\prime })\times \mathbf{E}^{\prime }$, respectively, in $\mathbf{f}%
_{S}^{\prime }$. The other two terms are: $(-\gamma _{S}/c^{4})\varepsilon
_{\alpha \mu \sigma \rho }\varepsilon ^{\alpha \beta \gamma \delta }(B_{\nu
}v_{\beta }-B_{\beta }v_{\nu })B^{\rho }v^{\sigma }d_{\gamma }u_{\delta }$,
and $(\delta _{Z}^{2}/c^{4}\gamma _{S})\varepsilon _{\alpha \mu \sigma \rho
}\varepsilon ^{\alpha \beta \gamma \delta }(E_{\nu }v_{\beta }-E_{\beta
}v_{\nu })B^{\rho }v^{\sigma }m_{\gamma }u_{\delta }$. They would correspond
to $\gamma _{S}(\mathbf{d}^{\prime }\times \mathbf{B}^{\prime })\times
\mathbf{B}^{\prime }$ and $(\delta _{Z}^{2}/\gamma _{S})(\mathbf{m}^{\prime
}\times \mathbf{E}^{\prime })\times \mathbf{B}^{\prime }$, respectively.
Such terms do not exist in $\mathbf{f}_{S}^{\prime }$, Eq. (26) in [1]. To
see why there is this difference we have repeated the derivation of Eq. (26)
in the same way as in [1], i.e., assuming that $\mathbf{\mu }=\gamma _{S}%
\mathbf{S}$ and $\mathbf{d}=\delta _{S}\mathbf{S}$ and therefore using only
the commutator $[S^{i},S^{j}]$. That calculation revealed that the result
quoted in [1] is not unique. Instead of the choice $\delta _{S}(\mathbf{m}%
^{\prime }\times \mathbf{B}^{\prime })\times \mathbf{B}^{\prime }$ one can
choose another equally well defined possibility $\gamma _{S}(\mathbf{d}%
^{\prime }\times \mathbf{B}^{\prime })\times \mathbf{B}^{\prime }$.
Similarly instead of $\delta _{S}(\mathbf{d}^{\prime }\times \mathbf{E}%
^{\prime })\times \mathbf{B}^{\prime }$ one can take $(\delta
_{S}^{2}/\gamma _{S})(\mathbf{m}^{\prime }\times \mathbf{E}^{\prime })\times
\mathbf{B}^{\prime }$. On the other hand our results are unique and they
correspond to the last two expressions, which are not chosen in [1]. This
means that even the expression for $\mathbf{f}_{S}^{\prime }$, Eq. (26) in
[1], has to be changed according to our results.\bigskip

\noindent \textbf{7. Comparison with experiments }\bigskip

Furthermore, in [1], different experiments are discussed and compared with
the expressions for $a^{\prime \mu }$ Eq. (16), $\mathbf{f}^{\prime }$ Eq.
(25) and $\mathbf{f}_{S}^{\prime }$ Eq. (26). However, as already said, such
comparison is not valid since these equations refer to the $K^{\prime }$
frame, while the experiments are performed in $K$. Hence, it is not true, as
declared in [1], that, e.g., the term $(-\gamma _{S}/c^{2})(\mathbf{m}%
^{\prime }\times \mathbf{B}^{\prime })\times \mathbf{E}^{\prime }$ ``may be
experimentally detectable.'' (According to [1] such possibility is explained
in Ref. [15] in [1]. The objections to that explanation will not be
considered here.) The same thing happens with the discussion of all other
experiments that is presented in [1]. Note that the AT of $\mathbf{E}$ and $%
\mathbf{B}$, and not the LT of $E^{\mu }e_{\mu }$ and $B^{\mu }e_{\mu }$,
are usually employed in the discussion of the experiments in [1].

Here, we shall show that $G_{\mu \nu }$ and the quantum force are not zero
in two interferometric experiments that have been proposed in [1]. In the
first experiment it is taken that the interfering particles have magnetic
moment, but no electric charge and zero or negligible electric dipole
moment, as in a neutron interferometer. The entire interferometer is
subjected to a homogeneous and time independent electric field $\mathbf{E}$
that is parallel to a pair of arms of the interferometer. From the
expression for $G_{\mu \nu }^{\prime }$ and $\mathbf{f}_{S}^{\prime }$, Eq.
(26) in [1], it is visible that both quantities are zero in the considered
experiment, as argued in [1]. However, here, in contrast to [1], this
experiment will be examined directly in $K$, i.e., $K$ is chosen to be the $%
e_{0}$-frame. First, when $B^{\mu }=d^{\mu }=0$ then there are only two
terms in $G_{\mu \nu }$ and consequently in $f_{S}^{\lambda }e_{\lambda }$,
which are different from zero. (Remember that $f_{S}^{\lambda }e_{\lambda }$
is formed taking the expectation value of $G_{\mu \nu }$ and multiplying it
by $u^{\nu }g^{\lambda \mu }e_{\lambda }$.) The general forms of these terms
(for $G_{\mu \nu }$) are: $(\delta _{Z}^{2}/c^{5}\gamma _{S})\varepsilon
^{\alpha \beta \gamma \delta }(E_{\mu }E_{\beta }v_{\nu }v_{\alpha }+E_{\nu
}E_{\alpha }v_{\mu }v_{\beta })m_{\gamma }u_{\delta }$ and $(\gamma
_{S}/4c^{7})\varepsilon ^{\alpha \beta \gamma \delta }\varepsilon _{\beta
\nu \sigma \rho }\varepsilon _{\alpha \mu \zeta \eta }(E^{\sigma }v^{\rho
}-E^{\rho }v^{\sigma })(E^{\zeta }v^{\eta }-E^{\eta }v^{\zeta })m_{\gamma
}u_{\delta }$. In $K$ it holds that $v^{\mu }=(c,0,0,0)$ and hence $E^{0}=0$%
. Inserting these values into the quoted terms one can see that $%
f_{S}^{0}e_{0}\neq 0$ (from the first term) and $f_{S}^{i}e_{i}\neq 0$ (from
both terms). For the lack of space these expressions will not be written.
The same would happen for the dual experiment that is proposed in [1]. This
consideration proves our assertion from the beginning of this paragraph.
This means that the phase shifts in these experiments are not due to
force-free interaction of the dipole, i.e., they are not topological phase
shifts. The same would happen for the Aharonov-Casher and the R\"{o}ntgen
phase shifts; the only difference is that for them the fields are not
homogeneous. It is interesting to note, as shown in [4], that even the
classical 4-force $(1/2)D^{ab}\partial ^{c}F_{ab}$ is not zero in the case
of the Aharonov-Casher and the R\"{o}ntgen effects. Also, it is worth noting
that the first model from [1] can be treated in a similar way using the
commutation relations for the 4D quantities. The results from [1] are
already used in several papers, e.g., [13]. Our discussion applies in the
same measure to their results.\bigskip

\noindent \textbf{8. Conclusions }\bigskip

In conclusion, here it is revealed that a unified and fully relativistic
treatment of the interaction from [1] can be achieved when in all steps of
the calculation only the 4D geometric quantities are used and not the 3D
quantities. It is expected that the obtained commutation relations for the
angular momentums (\ref{i}), and the analogous ones for the intrinsic
angular momentums (\ref{sz}), and those for the dipole moments (\ref{1}),
will greatly influence the existing quantum field theories. Finally, the
result that the quantum 4-force $f_{S}^{\lambda }e_{\lambda }$ for the
second model from [1] (and similarly $f^{\lambda }e_{\lambda }$ for the
first model) is different from zero will significantly change the usual
explanations of the quantum phase shifts in, e.g., the neutron
interferometry, the Aharonov-Casher and the R\"{o}ntgen effects.\bigskip

\noindent \textbf{References\bigskip }

\noindent \lbrack 1] J. Anandan, Phys. Rev. Lett. 85 (2000) 1354.

\noindent \lbrack 2] T. Ivezi\'{c}, Phys. Rev. Lett. 98 (2007) 108901.

\noindent \lbrack 3] T. Ivezi\'{c}, physics/0703139.

\noindent \lbrack 4] T. Ivezi\'{c}, Phys. Rev. Lett. 98 (2007) 158901.

\noindent \lbrack 5] A. Einstein, Ann. Physik 17 (1905) 891, tr. by W.
Perrett and G.B. Jeffery,

in The Principle of Relativity, Dover, New York, 1952.

\noindent \lbrack 6] T. Ivezi\'{c}, Found. Phys. Lett. 18 (2005) 401.

\noindent \lbrack 7] F.J.M. Farley et al., Phys. Rev. Lett. 93 (2004) 052001.

\noindent \lbrack 8] T. Ivezi\'{c}, Found. Phys. 33 (2003) 1339\textbf{; }T.
Ivezi\'{c}, Found. Phys. Lett. 18

(2005) 301; T. Ivezi\'{c}, Found. Phys. 35 (2005) 1585.

\noindent \lbrack 9] T. Ivezi\'{c}, Found. Phys. 36 (2006) 1511.

\noindent \lbrack 10] J.D. Jackson, Classical Electrodynamics, 3rd ed.
Wiley, New York, 1998.

\noindent \lbrack 11] J. D. Jackson, Am. J. Phys. 72 (2004) 1484.

\noindent \lbrack 12] S. Weinberg, The Quantum Theory of Fields, Volume I,
Foundations, Cambridge University, Cambridge, 1999.

\noindent \lbrack 13] C. Furtado and C. A. de Lima Ribeiro, Phys. Rev. A 69
(2004) 064104.

\end{document}